\documentclass{iaus}
\usepackage{graphicx}
\usepackage{sidecap}   

\newcommand{\kms}{km~s$^{-1}$\,}
\newcommand {\sm}{\rm\,M$_\odot$}

\renewcommand{\theequation}{\arabic{equation}}
\title[The mass content of the Scl dSph] 
{The mass content of the Sculptor dwarf spheroidal galaxy}

\author[G.Battaglia et al.]   
{G.Battaglia$^{1,2}$,
 A.Helmi$^1$, E.Tolstoy$^1$, \and M.Irwin$^3$}

\affiliation{$^1$Kapteyn Astronomical Institute, University of Groningen, P.O.Box 800, the Netherlands \\[\affilskip]
$^2$European Southern Observatory, K. Schwarzschild-Str. 2, 85748 Garching, Germany\\ email: {\tt gbattagl@eso.org} \\[\affilskip] $^3$Institute of Astronomy, Madingley Road, Cambridge CB03 0HA, UK}

\pubyear{2008}
\volume{254}  
\jname{The Galaxy Disk in Cosmological Context}
\editors{J.~Andersen, J.~Bland-Hawthorn \& B.~Nordstr\"{o}m, eds}
\begin{document}

\maketitle

\begin{abstract}
We present a new determination of the mass content of the Sculptor dwarf spheroidal galaxy, 
based on a novel approach which takes into account the two distinct stellar populations 
present in this galaxy. This method helps to partially break the well-known mass-anisotropy 
degeneracy present in the modelling of pressure-supported stellar systems.
\end{abstract}

\firstsection 
\section{Introduction}
The determination of the mass content of galaxies is a fundamental step 
for our understanding of their formation and evolution. This is particularly true for 
small systems such as dwarf spheroidal galaxies (dSphs), whose total mass 
can determine their destiny as being fragile objects, strongly perturbed by supernovae explosions and/or 
interactions with the environment, or mostly undisturbed objects. 

Part of the interest in determining the masses of dSphs has surely been driven by the 
fact that these galaxies, with mass-to-light (M/L) ratios up to 100s, appear to be the most 
dark matter (DM) dominated objects known to-date. 
This makes them potentially good testing grounds for different DM theories of galaxy formation; 
however at the moment it is still unclear if dSphs inhabit 
cusped DM profiles, as predicted by cold DM theories of galaxy formation, 
or cored profiles, typical of warmer kinds of DM. This distinction is 
often hampered by the degeneracies which are intrinsic to the method generally used to determine 
the mass of pressure-supported stellar systems, i.e. a Jeans analysis of the line-of-sight (l.o.s.)
velocity dispersion obtained considering all stars as a single
component embedded in an extended DM halo. This analysis is subject to
the well known degeneracy between the mass distribution and the
orbital motions of the individual stars in the system (mass-anisotropy
degeneracy). Recent observations have shown that some dSphs 
host multiple stellar populations which have distinct spatial distribution, metallicity 
and kinematics. The presence of multiple stellar populations requires a modification in the
way these systems are dynamically modelled to derive their mass content.

Here we describe the results of our determination of 
 the mass content of the Sculptor (Scl) dSph by adopting a more detailed kinematic modelling than 
 the traditional one, i.e. by 
considering Scl as two-(stellar) components embedded in an extended DM halo. 
We first briefly describe and discuss the observational properties of Scl derived from our work 
(\cite[Tolstoy et al. 2004]{tolstoy2004}; \cite[Battaglia 2007]{battaglia2007}; \cite[Battaglia et al. 2008]{battaglia2008}).
We then move to 
the results of the two-component modelling; we show 
a clear example of the mass-anisotropy present when modelling Scl as a one-stellar component system, 
and finally discuss our results. 
For a more detailed description we refer to \cite[Battaglia (2007)]{battaglia2007} and 
\cite[Battaglia et al. (2008)]{battaglia2008}.

\section{Observed properties of the Sculptor dSph}
We acquired extended photometry from the ESO/2.2m WFI and VLT/FLAMES 
spectra (R$\sim$6500) in the Ca\,{\small II}\, triplet region for hundreds of individual red giant branch (RGB) 
stars in Scl, out to its nominal 
tidal radius (\cite[Tolstoy et al. 2004]{tolstoy2004}: 308 probable members; 
\cite[Battaglia et al. 2008]{battaglia2008}: 470 probable members). 
We used these data to study the large scale kinematics and metallicity properties 
of Scl, and the spatial distribution of its stellar populations.

The combined information from the accurate line-of-sight velocities ($\pm$2 \kms) and 
metallicities ($\pm$0.15 dex) of RGB stars allowed us to unveil the presence of 
two distinct stellar populations with different spatial distribution, metallicity and kinematics: the 
metal rich (MR) stars ([Fe/H]$>-1.5$) are more centrally concentrated and 
show colder kinematics than the metal poor (MP) stars ([Fe/H]$<-1.7$), as can be 
seen from the line-of-sight (l.o.s.) velocity dispersion profiles in Fig.~\ref{fig:mrmp}. 
The MR and MP RGB stars appear to trace,  respectively, the red and blue horizontal branch (RHB, BHB) stars. 
The surface number density profile of RGB stars, representative of the overall stellar population of Scl, 
is well approximated by a two-component fit, where each component is given by
the rescaled best-fitting profile derived from our photometry for the
RHB and BHB stars. In contrast, a single profile (be it a Plummer, Sersic or King) 
does not appear to be a good approximation for the surface number density of RGB stars. 
Multiple stellar populations are known to exist also in other dSphs, such as Fornax 
(\cite[Battaglia et al. 2006]{battaglia2006}) and Canes Venatici (\cite[Ibata et al. 2006]{ibata2006}), 
and might be a common feature to this class of objects.

The large spatial coverage and statistics of this spectroscopic data-set has allowed the discovery 
of another interesting feature, i.e. a velocity gradient of 7.6 \kms deg$^{-1}$ in the 
Galactic standard of rest (GSR) \footnote{We use the velocities in the GSR
frame to avoid spurious gradients introduced by the component of the
Sun and Local Standard of Rest motion's along the l.o.s.\ to Scl.}
along the projected major axis of Scl. This gradient is likely to due intrinsic rotation, 
and not to tidal disruption (see \cite[Battaglia 2007]{battaglia2007} and 
\cite[Battaglia et al. 2008]{battaglia2008} 
for discussion on this point), and 
this is the first time that statistically significant rotation is found in a dSph.

Among the other Milky Way (MW) dSphs, a significant velocity 
gradient is detected in Carina (\cite[Mu\~{n}oz et al. 2006]{munoz2006car}) and a mild one in Leo~I 
(\cite[Mateo et al. 2008]{mateo2008}), but these are interpreted 
as the result of tidal interaction with the MW. However, velocity gradients are not 
exclusive of the dSphs satellites of larger galaxies, but are present 
also in some of the isolated Local Group dSphs, such as Cetus (\cite[Lewis et al. 2007]{lewis2007}) 
and Tucana (Fraternali et al., in preparation), which are unlikely to have ever interacted 
with either the MW or M31 and therefore suffered tidal disturbance. 
This might 
point to rotation as an intrinsic property of this class of objects and provide insights 
into the mechanisms that shaped their evolution. This might for instance favour scenarios in which 
the progenitors of dSphs were rotating systems which were transformed into prevalently pressure 
supported object by tidal stirring from their host galaxy (e.g., \cite[Mayer et al. 2001]{mayer2001}). 
It would therefore be interesting to compare the frequency of 
occurrence of rotation among dSphs and other generally more isolated systems such as 
dwarf irregular galaxies and transition types. It is possible that more velocity gradients 
will be discovered in MW dSphs now that larger and more spatially extended 
velocity surveys are becoming available.

In the analysis presented here we always use rotation-subtracted GSR 
velocities.

\section{The Jeans analysis}
DSphs generally have low ellipticities and are presumably prevalently pressure supported, and therefore 
are commonly considered as spherical and with no streaming motions\footnote{We checked that the assumptions of 
sphericity and absence of streaming motions have a
negligible effect on the results.}. 
Then, if the system is in dynamical equilibrium,
 the l.o.s.\ velocity dispersion predicted by the Jeans equation is (from 
\cite[Binney \& Mamon 1982]{binney1982})

\setcounter{equation}{0}
\begin{equation}
\label{eq:jeans_binneymamon1982}
\sigma_{\rm los}^2(R)= \frac{2}{\Sigma_*(R)} \int_R^{\infty} \frac{\rho_*(r) \sigma_{r,*}^2 \ r}{\sqrt{r^2-R^2}}
(1 - \beta \frac{R^2}{r^2} ) dr
\end{equation}
where $R$ is the projected radius (on the sky) and $r$ is the 3D radius. 
The l.o.s.\ velocity dispersion depends on:
 the mass surface density $\Sigma_*(R)$ and mass density $\rho_*(r)$
of the tracer, which in our case are the MR and the MP RGB stars;
the tracer velocity anisotropy $\beta$, defined as $\beta = 1 -\  \sigma_{\theta}^2 / \sigma_r^2$, 
and its radial velocity dispersion $\sigma_{r,*}$, 
which depends on the total mass distribution (for dSphs the contribution due to 
stellar mass is negligible). 
A unique solution for the
mass profile can be determined from this equation if both the mass distribution of the tracer and
$\beta(r)$ are known, although this solution is not guaranteed to produce a
phase-space distribution function that is positive everywhere. However, 
proper motions for individual stars in the system would be necessary to derive $\beta(r)$ and this is not feasible yet. 
This causes the well-known mass-anisotropy
degeneracy, which prevents a distinction amongst different DM
profiles. Examples of this are present in the literature: 
the observed l.o.s. dispersion profiles of dSphs are both compatible with 
cored DM models, assuming that $\beta=0$ (e.g., \cite[Gilmore et al. 2007]{gilmore2007}),  
and with cusped profiles,  allowing for mildly tangential, constant with radius, $\beta$ 
(e.g. \cite[Walker et al. 2007]{walker2007}). 
It is therefore advisable to explore several hypotheses for the behaviour of $\beta(r)$ and 
for the mass distribution when carrying out a Jeans analysis (see below).

In the following we explore the possibility that DM follows a cored (pseudo-isothermal 
sphere) or a cuspy distribution (NFW profile); 
we also explore two hypotheses for the behaviour of $\beta(r)$: constant with radius, and 
following an Osipkov-Merritt (OM) model (isotropic in the central parts and radial in the outskirts). 
The spatial distributions of the tracers are derived from our photometric data.

\begin{figure}[!ht]
\begin{center}
\includegraphics[width=0.6\textwidth]{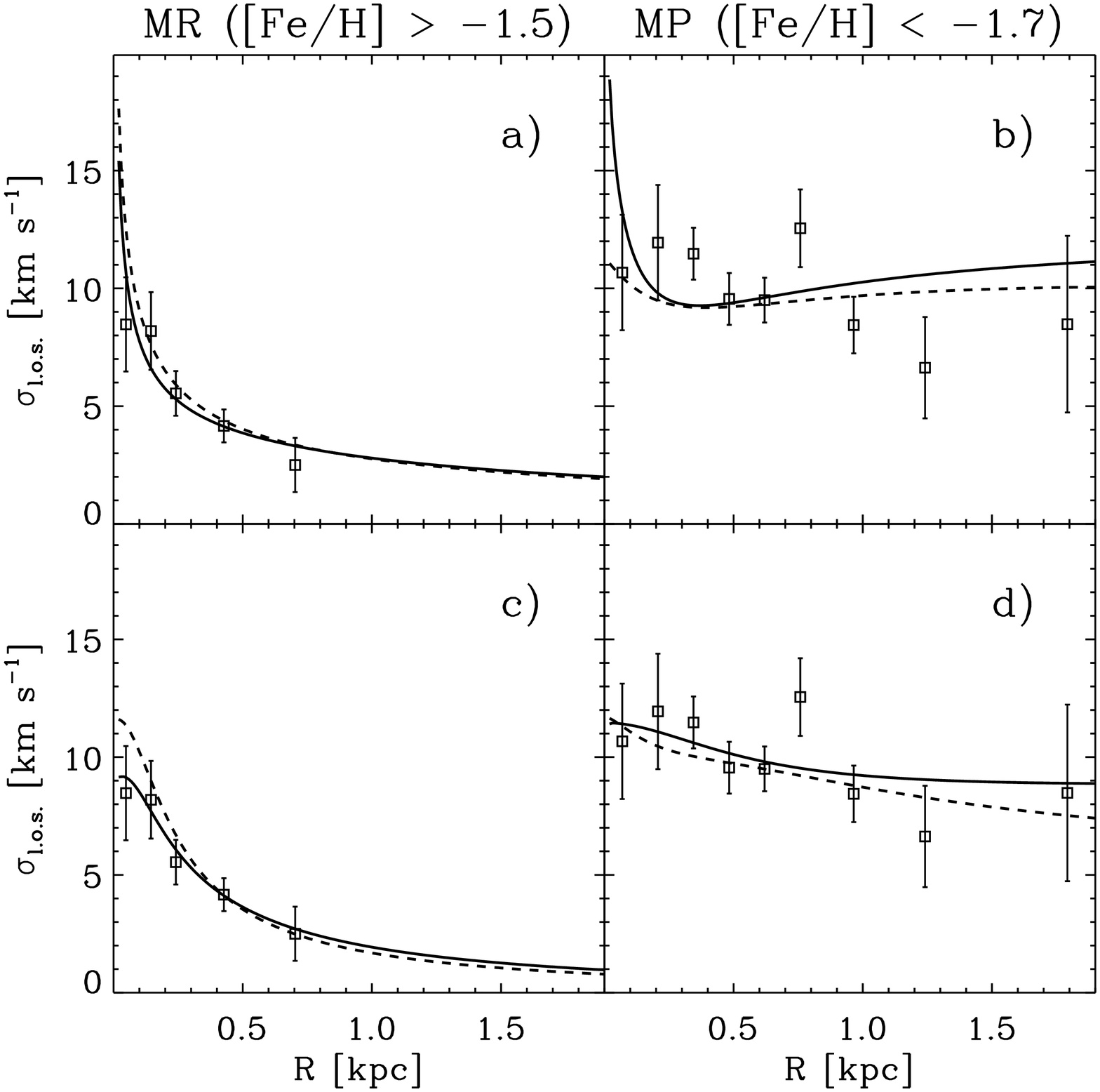} 
\vspace*{-0.4 cm}
\caption{L.o.s.\ velocity dispersion profile (squares with errorbars), from rotation-subtracted GSR velocities, 
for the MR (left) and MP (right) RGB stars in Scl. The lines show the best-fitting 
pseudo-isothermal sphere (solid) and NFW model (dashed) in the hypothesis of $\beta=const$  
(a,b) and $\beta=\beta_{\rm OM}$ (c,d). The models have reduced $\chi^2 \sim 1.6$ (a,b), 0.6 (c) and 1 (d).}
   \label{fig:mrmp}
\end{center}
\end{figure}

{\underline{\it Two-components modelling}}  
Here we show the results of modelling Scl as a two-(stellar) component system
embedded in a DM halo. Eq.~\ref{eq:jeans_binneymamon1982} is applied separately to the MR and MP stars, 
assuming that they move in the same DM potential. 
We explore a range of core radii $r_c$ for the pseudo-isothermal
sphere ($r_c\!=\!0.001, 0.05, 0.1, 0.5, 1$ kpc) and  of
concentrations $c$ for the NFW profile ($c\!=\!20, 25, 30, 35$). By fixing
these, each mass model has two free parameters left: the anisotropy and the DM halo mass
(enclosed within the last measured point for the isothermal sphere, at
1.3 deg $=$ 1.8 kpc assuming a distance to Scl of 79 kpc, see \cite[Mateo 1998]{Mateo1998}; 
and the virial mass in the case of the NFW model).  We
compute a $\chi^2$ for the MR and MP components separately ($\chi_{\rm
MR}^2$ and $\chi_{\rm MP}^2$, respectively) by comparing the various
models to the data. The best-fit is obtained by minimising
the sum $\chi^2= \chi_{\rm MR}^2 + \chi_{\rm MP}^2$.

Figure~\ref{fig:mrmp}a,b shows the models with constant anisotropy, which can be excluded as in this 
case neither a cored or a cusped model is a good representation of the data 
($\chi_{\rm min}^2 \sim 17$). The situation is different in the
hypothesis of an OM velocity anisotropy as shown in Fig.~\ref{fig:mrmp}c,d). We find that a
pseudo-isothermal sphere with a relatively large core ($r_c=$0.5 kpc, $M(< 1.8 \mathop{\rm
kpc}) = 3.4 \pm 0.7 \times 10^8$ \sm) gives an excellent description of the data ($\chi_{\rm min}^2= 6.9$). 
Also an NFW model is statistically consistent with
the data ($c=20$, virial mass $M_v= 2.2_{-0.7}^{+1.0} \times 10^9$
\sm, $\chi_{\rm min}^2= 10.8$), but tends to over-predict the
central values of the MR velocity dispersion. The mass within 
the last measured point ($\sim 2.4_{-0.7}^{+1.1} \times 10^8$ \sm) 
is consistent with the mass predicted by the best-fitting
isothermal sphere model. 

\begin{figure}[!h]
\begin{center}
\includegraphics[width=0.9\textwidth]{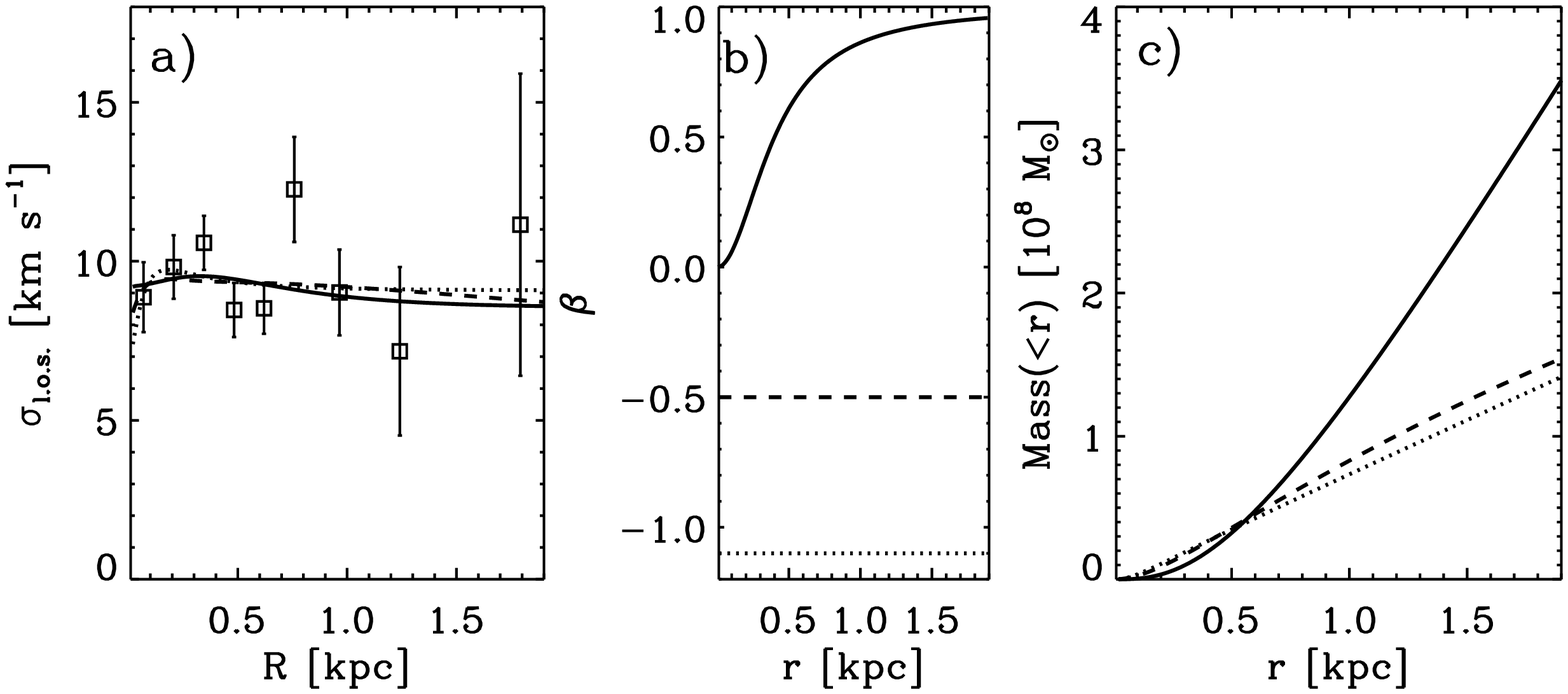} 
\caption{a) Observed l.o.s. velocity dispersion profile, from rotation-subtracted GSR velocities,  
for all RGB stars in Scl (squares with error-bars), with 
overlaid the best-fits for a cored DM distribution (solid line: $\beta=\beta_{\rm OM}$ and $r_c=$0.5 kpc; 
dotted line: $\beta=$ const.and $r_c=$0.05 kpc), and an NFW DM distribution (dashed line: $\beta=$ const. and $c=35$). 
Panel b) and c) show the velocity anisotropy and mass distribution corresponding 
to the models in panel a). Note that the model with $r_c=$0.05 kpc is almost equivalent to a cusp.}
   \label{fig:all}
\end{center}
\end{figure}

{\underline{\it One-component modelling}} 
Figure~\ref{fig:all}  shows a  clear example of the 
mass-anisotropy degeneracy when modelling Scl with the traditional one-component analysis. 
Just by changing the assumption on $\beta$ ($=$const. or $=\beta_{\rm OM}$), 
cored profiles with very different core radii and $M(< 1.8 \mathop{\rm kpc})$ give 
excellent fit to the data. An excellent fit is also an NFW profile, 
in the hypothesis of $\beta =$ const. 
These models are practically indistinguishable and all good ($\chi^2\sim 8$, 
reduced $\chi^2\sim 1.1$). 
No DM model can be favoured nor hypotheses on $\beta$. 

\section{Discussion and conclusions}
The combined information of velocity and metallicity from a large sample of FLAMES spectra 
of individual stars in Scl has allowed us to separate the different kinematics of the two stellar 
populations in Scl, and to carry out a more detailed kinematic modelling which allows us 
to partially break the mass-anisotropy degeneracy.

Under the hypotheses explored for the behaviour of the velocity anisotropy $\beta(r)$, 
the two-component modelling indicates that a cored profile is slightly favoured over a cusped profile, 
as the latter tends to over-predict the central values (at $R \lesssim$ 0.1 kpc) of the MR velocity dispersion. 
It could be argued that the relatively large central dispersion predicted for the MR stars 
from the cusped profile could be lowered by allowing the velocity anisotropy of the MR population 
to be tangential at $R \lesssim$ 0.1 kpc. 
However it should be noted that in order to reproduce the rapid decline of the MR velocity dispersion profile, the MR 
$\beta(r)$ needs to become very radial already on the scale of 0.3-0.4 kpc, given the observed spatial distribution of MR stars. 
Observational determinations 
of $\beta$, either from direct measurement or from distribution functions, would be the most desirable 
(and definitive) solution to this issue. 

\cite[Strigari et al. (2007)]{strigari2007} showed that, 
independently of $\beta(r)$ and the total mass distribution of dSphs, the 
mass enclosed within 0.6 kpc (M$_{\rm 0.6}$) can be derived more accurately than 
at any other distance from the centre. However, given that the evolution of dSphs is most likely to be governed by 
their {\it total} mass and that dynamical 
analyses provide only the {\it mass enclosed within the last measured point}, 
it is important to have reliable determinations of the mass enclosed 
as farther out as possible. Our analysis shows that, independently of the 
adopted mass profile, the best fits from our two-component modelling yield consistent values 
for the mass enclosed within the last measured point, at 1.8kpc.

\begin{figure}[!ht]
\begin{center}
\includegraphics[width=0.5\textwidth]{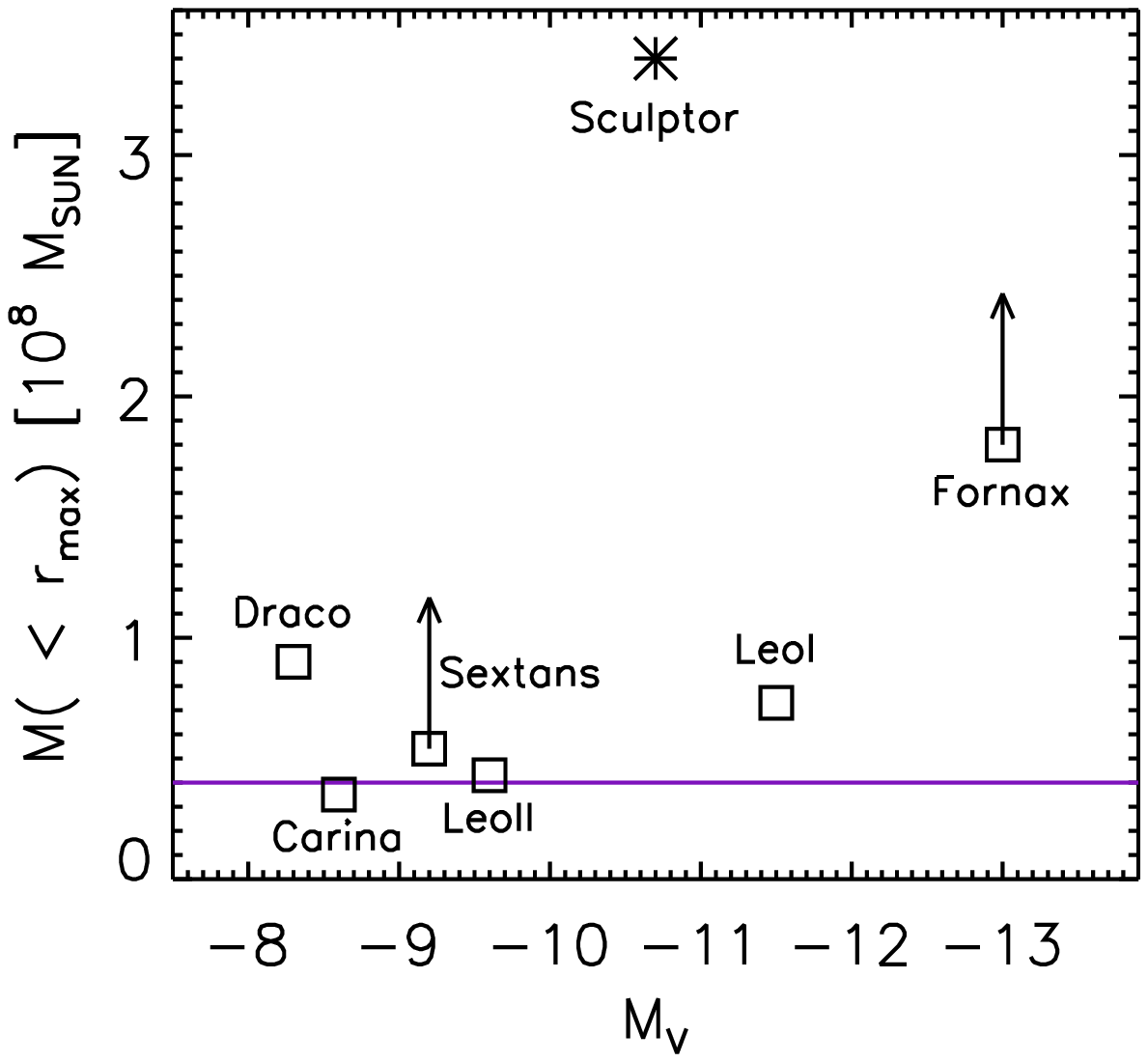} 
\includegraphics[width=0.45\textwidth]{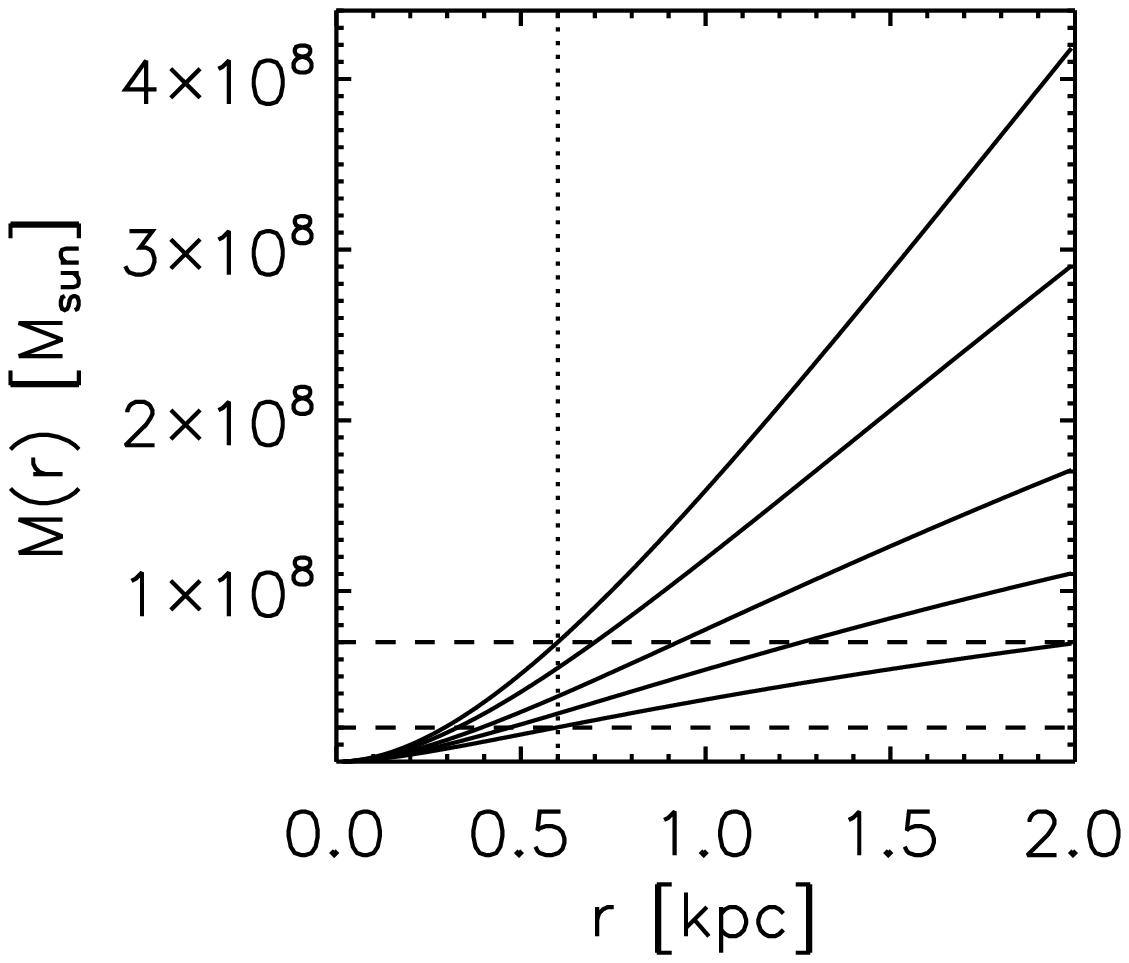}
\caption{Left: Mass within the last measured point ($r_{\rm max}$) versus absolute $V$ magnitude, M$_V$, for MW dSphs. 
The masses are from \cite[Walker et al. 2007]{walker2007} (squares), 
except for Scl which is from our two-components modelling (asterisk); 
M$_V$ is from \cite[Irwin \& Hatzidimitriou (1995)]{ih1995}. The arrows indicate that the masses measured 
for Sextans and Fornax are likely to increase given that these objects have been surveyed just to 0.3 and 0.6 tidal radii. 
The solid line indicates a mass of 4$\times 10^7$ \sm. 
Right: Mass enclosed within $r$ for NFW halos of virial masses 2,4,8,20,40$\times 10^8$ \sm (solid lines), i.e. the range found by 
\cite[Walker et al. (2007)]{walker2007} to fit the observed l.o.s. velocity dispersion profile of their sample of 7 dSphs. 
Concentrations 
are derived using the formula in \cite[Koch et al. (2007)]{koch2007}, extrapolated from \cite[Jing \& Suto (2000)]{jing2000}. 
The dashed lines enclose masses between 2-7$\times 10^7$ \sm; the dotted line is placed at a radius of 0.6 kpc.
}\label{fig:mass}
\end{center}
\end{figure}

Finally we would like to comment on the issue that dwarf galaxies might be embedded in haloes of similar mass 
($\sim 4 \times 10^7$ \sm) as has been claimed in the literature. 
Figure~\ref{fig:mass}(left) shows a revisited version of 
the mass enclosed within the last measured point versus absolute magnitude plot. 
The masses plotted are the ones derived in \cite[Walker et al. (2007)]{walker2007} for a sample of dSphs, 
because they are derived from the most extended velocity samples and from a homogeneous analysis. 
For Scl we use our own determination, which is the most accurate so far. The figure shows that 
the DM mass of Scl is now much larger than 
the typical DM mass derived so far for the other dSphs. Also Fornax deviates considerably from the relation. 
In the plot shown here, the masses of Fornax and Sextans were derived from data 
extending out to $\sim$0.6 and 0.3 tidal radii, respectively. 
Given that Fornax and Sextans have been recently mapped much farther out 
(Fornax out to $\sim$ 1 tidal radius: Battaglia et al. 2006; Sextans out to $\sim$ 0.8 tidal radius: 
Battaglia et al. in preparation), 
their revised masses are likely to increase quite significantly, and introduce even larger dispersion in this plot. 
In some cases, M$_{\rm 0.6}$ has been used in this kind of analysis, instead than the mass enclosed within 
the last measured point. However, 
Fig.~\ref{fig:mass}(right) shows that M$_{\rm 0.6}$ is very insensitive to the total mass of the dSph. 
In fact, NFW haloes 
of virial masses between 2-40 $\times 10^8$ \sm (i.e. in the range found by \cite[Walker et al. 2007]{walker2007} 
to fit their samples of dSphs) give remarkably similar values of M$_{\rm 0.6}$, probably because 
at such a small distance there is not much sensitivity to the total mass. Therefore there is no indication 
of a common mass scale amongst the dSphs in the Local Group, while perhaps one may draw the conclusion of the 
existence of a minimum mass for the formation of dSphs.


\begin{thebibliography}{}

\bibitem[Battaglia et al.(2006)]{battaglia2006} Battaglia, G., et al. 2006, \textit{A\&A}, 459, 423

\bibitem[Battaglia (2007)]{battaglia2007} Battaglia, G. 2007, PhD thesis, Univ. Groningen, The Netherlands, http://irs.ub.rug.nl/ppn/304002712

\bibitem[Battaglia et al. (2008)]{battaglia2008} Battaglia, G., et al. 2008, \textit{ApJ} (Letters), 681, L13 

\bibitem[Binney \& Mamon (1982)]{binney1982} Binney, J. \& Mamon, G.~A. 1982, \textit{MNRAS}, 200, 361

\bibitem[Gilmore et al.(2007)]{gilmore2007} Gilmore, G., et al. 2007, \textit{ApJ}, 663, 948

\bibitem[Ibata et al.(2006)]{ibata2006} Ibata, R., Chapman, S., Irwin, M., Lewis, G., \& Martin, N. 2006, \textit{MNRAS} (Letters), 373, L70

\bibitem[Irwin and Hatzidimitriou (1995)]{ih1995}  Irwin, M.~J., \& Hatzidimitriou, D. 1995, \textit{MNRAS}, 277, 1354

\bibitem[Jing \& Suto (2000)]{jing2000} Jing, Y.~P., \& Suto, Y. 2000, \textit{ApJ} (Letters), 529, L69

\bibitem[Koch et al. (2007)]{koch2007} Koch, A. et al. 2007, \textit{ApJ}, 657, 241

\bibitem[Lewis et al.(2007)]{lewis2007} Lewis et al. 2007, \textit{MNRAS}, 1364, 1370

\bibitem[Mateo (1998)]{Mateo1998} Mateo, M.~L. 1998, \textit{ARAA}, 36, 435

\bibitem[Mateo et al. (2008)]{mateo2008} Mateo, M., Olszewski, E.~W., Walker, M.~G. 2008, \textit{ApJ}, 675, 201

\bibitem[Mayer et al.(2001)]{mayer2001} Mayer, L., et al. 2001, \textit{ApJ}, 559, 754

\bibitem[Mu\~{n}oz et al.(2006)]{munoz2006car} Mu\~{n}oz, R.~R. et al. 2006, \textit{ApJ}, 649, 201

\bibitem[Strigari et al.(2007)]{strigari2007} Strigari, L.~E., et al. 2007, \textit{ApJ}, 669, 676

\bibitem[Tolstoy et al.(2004)]{tolstoy2004} Tolstoy, E., et al. 2004, \textit{ApJ} (Letters), 617, L119

 \bibitem[Walker et al.(2007)]{walker2007} Walker, M.~G., et al. 2007, \textit{ApJ} (Letters), 667, L53

\end{thebibliography}
\end{document}